%% file: main.tex
\documentclass[preprint,1p]{elsarticle}
\usepackage{amsmath}

\usepackage{graphicx}
\usepackage{amssymb, amsmath}
\usepackage{bm}
\usepackage[ruled,lined,algonl]{algorithm2e}
\usepackage{dcolumn}
\usepackage{amsmath}
\usepackage{amsfonts}
\usepackage{mathtools}
\usepackage{mathrsfs}
\usepackage{amsthm}

\input{macro.tex}
\begin{document}

\begin{frontmatter}



\title{Analyzing the Dual Space of the Saturated Ideal of a Regular Set and the Local Multiplicities of its Zeros}


\author[a]{Xiaoliang Li}
\ead{xiaoliangbuaa@gmail.com}

\author[b,c]{Wei Niu\corref{corr}}
\ead{wei.niu@buaa.edu.cn}

\cortext[corr]{Corresponding author.
Postal address: Ecole Centrale de P\'ekin, Beihang University, Beijing, 100191, China}

\address[a]{School of Finance and Trade, City College of Dongguan University of Technology, Dongguan 523419, China}

\address[b]{Beijing Advanced Innovation Center for Big Data and Brain Computing, Beihang University, Beijing, 100191, China}
\address[c]{Ecole Centrale de P\'ekin, Beihang University, Beijing, 100191, China}


\begin{abstract}
In this paper, we are concerned with the problem of counting the multiplicities of a zero-dimensional
regular set's zeros. We generalize the squarefree decomposition of univariate polynomials to the so-called pseudo
squarefree decomposition of multivariate polynomials,
and then propose an algorithm for decomposing a regular set into a finite number of simple sets.
From the output of this algorithm,
{the multiplicities of zeros could be directly read out, and the real solution isolation with multiplicity
 can also be easily produced.}
As a main theoretical result of this paper, we analyze the structure of dual space of the saturated ideal generated by a simple set as well as a regular set. Experiments with a preliminary implementation show the efficiency of our method.
\end{abstract}


\end{frontmatter}

\section{Introduction}

Polynomial equations are widely used in science and engineering to describe
various problems. The multiplicities of the solutions are crucial
characteristics, which help us to intensively understand the algebraic structure behind equations.

The study of multiplicities at solutions of polynomial equations may be traced back to
the foundation of algebraic geometry. After that, researchers did many remarkable work on this topic.
Based on the dual space theory, Marinari and others \cite{m95g} proposed an algorithm for computing the multiplicity.
Furthermore, the computation of multiplicity structure could be reduced to solving eigenvalues
of the so-called multiplicity matrix, which is studied by M\"oller and Stetter \cite{m95m},
Stetter \cite{s96m} and others. Interested readers may find other relative literature given in \cite{d11m,d05c}.

\begin{example}\label{ex:intro}
Consider the univariate polynomial $F=x^5-x^3$ for example. It is easy to verify that $0$ is a zero of $F$ and
\begin{equation*}
 F'(0)=0,\quad F''(0)=0,\quad F^{(3)}(0)\neq 0.
\end{equation*}
It follows that the multiplicity of $0$ at $F$ is $3$. This is the fundamental idea of counting
the multiplicities of zeros using the dual space theory.
\end{example}

Triangular decomposition is one of main elimination approaches for
solving systems of multivariate polynomial equations.
The first well-known method of triangular decomposition is called the \emph{characteristic set} method,
which was proposed by Wu \cite{w86z,w86b} based on Ritt's work on differential ideals \cite{r50d}.
But the zero set of a characteristic set may be empty. To remedy this shortcoming,
Kalkbrener \cite{k93g}, Yang and Zhang \cite{y94s} introduced the notation of regular set.
The properties of regular sets and relative algorithms
have been intensively studied by many researchers such as Wang \cite{w00c}, Hubert \cite{h03n},
Lazard \cite{l91n} and Moreno Maza \cite{m00t}.
The reader may refer to
\cite{c11a,w93e,w98d,w00c,m00t,g91w,g91e,g93d,hsl21,mbl21,wdm20} and
references therein for other literature  on triangular decomposition of polynomial systems.

Li gave a method in \cite{l05m} to count the multiplicities of a zero-dimensional polynomial system's zeros
after decomposing the system into triangular sets. Motivated by his work, we consider a relative yet different problem: efficiently
counting the multiplicities of a regular set's zeros.
Our main idea is based on the observation that in Example \ref{ex:intro},
$F$ can be rewritten as $F=x^3(x^2-1)$
with $\gcd(x,x^2-1)=1$ and $x,x^2-1$ to be squarefree.
Then the multiplicity of $0$ can be directly read from the exponent of the factor $x-0$ in $F$.

In this paper, we extend the above philosophy to the multivariate case. To be exact,
we generalize the squarefree decomposition of univariate polynomials to the so-called pseudo
squarefree decomposition of multivariate polynomials,
and then propose an algorithm for computing the multiplicities of a regular set's zeros.
{The method proposed in this paper can also be used to produce the real solution
isolation with multiplicity as in \cite{z11r}. }As a main theoretical result of this paper,
we analyze the structure of dual space of the saturated ideal generated by a simple set as well as a regular set.

The rest of this paper is structured as follows. In section \ref{sec:pre},
basic notations and relative properties of multiplicity and triangular decomposition
are revisited. In section \ref{sec:psqf}, we introduce the pseudo squarefree decomposition of a multivariate
polynomial and give a feasible algorithm to compute it.
{Based on the pseudo squarefree decomposition, in section \ref{sec:mult} we propose the algorithm $\alg{Reg2Sim}$
with a regular set as its input, and the multiplicities of zeros can be easily obtained from the output.}
Section \ref{sec:exp} shows the efficiency of our approach with extensive experiments.

\section{Preliminaries}\label{sec:pre}

In what follows, we use $\p{x}$ to denote variables $x_1, \ldots, x_n$.
$\cnum[x_1, \ldots, x_n]$ or simply $\cnum[\xvar]$ represents the  polynomial ring
 with a fixed variable ordering $x_1 < \cdots < x_n$.

\subsection{Multiplicity}

In \cite{d11m,d05c}, Dayton and others proposed methods for computing
the multiplicity structure of zeros of a zero-dimensional polynomial system.
Their approach is based on the theory of dual space. In this section,
we revisit relative notations and theorems.

Let $\nnum=\{0,1,2,\ldots\}$. For any index array $\js=[j_1,\dots,j_r]\in\nnum^r$, we define the
differential operator
$$\partial_\js\equiv\partial_{j_1\cdots j_r}\equiv\frac{1}{j_1!\cdots j_r!}
\frac{\partial^{j_1+\cdots+j_r}}{\partial x_1^{j_1}\cdots\partial x_s^{j_r}}.$$

Let $\pp$ be a zero of the zero-dimensional ideal $\pset{I}\subseteq\cnum[\xvar]$.
For any $\partial_\js$, we can define a functional
$\partial_\js[\pp]: \cnum[\xvar]\rightarrow\cnum$, where $\partial_\js[\pp](F)=(\partial_\js F)(\pp)$ for $F\in\cnum[\xvar]$.
Any element of the vector space over $\cnum$  spanned by $\partial_\js[\pp]$ is called a \emph{differential functional} at $\pp$.
All differential functionals at  $\pp$ that vanish on $\pset{I}$ form a subspace
$$\mathbb{D}_{\pp}(\pset{I})\equiv\left\{\sum_{\js\in\nnum^r}c_\js\partial_\js[\pp]: c_\js\in\cnum,~\text{and}~
\sum_{\js\in\nnum^r}c_\js\partial_\js[\pp](F)=0 \text{~for all~} F\in\pset{I}\right\},$$
which is called the \emph{dual space} of $\pset{I}$ at $\pp$.

\begin{definition}[Local Multiplicity]\label{def:dual-mult}
Suppose that $\pset{I}$ ia a zero-dimensional ideal in $\cnum[\xvar]$,
{i.e.\ $\pset{I}$ has a finite number of complex zeros.}
Let $\pp$ be a zero of $\pset{I}$.
The dimension of the vector space $\mathbb{D}_{\pp}(\pset{I})$ is
 called the \emph{local multiplicity} {or \emph{multiplicity} for short} of $\pp$  in $\pset{I}$.
\end{definition}

Let $\pset{S}$ be a multiplicatively closed subset of $\kx$.
We use $\pset{S}^{-1}\pset{I}$ to denote the localization of the polynomial ideal $\pset{I}$ at $\pset{S}$, i.e.\
$\pset{S}^{-1}\pset{I}\equiv\{F/G:\,F\in\pset{I},~G\in\pset{S}\}$.

\begin{theorem}[{\cite{d05c}}]\label{def:inter-mult}
  Under the assumption of Definition \ref{def:dual-mult}, the \emph{local multiplicity}
  of $\pp$  in $\pset{I}$ equals to
  the dimension of the quotient ring $\pset{S}^{-1}\cnum[\xvar]/\pset{S}^{-1}\pset{I}$ as
  a vector space over $\cnum$, where $\pset{S}=\kx\setminus\pset{M}_\pp$ and $\pset{M}_\pp$ is the maximal ideal of $\pp$.
\end{theorem}

\subsection{Triangular Decomposition}

Let $F$ and $G$ be two polynomials in $\kx$.
{The variable of biggest index} appearing in $F$ is
called the \alert{leading variable} of $F$ and denoted by $\lv(F)$.
The leading coefficient of $F$, viewed as a univariate polynomial in $\lv(F)$, is called the
\emph{initial} of $F$ and denoted by $\ini(F)$.
Moreover, $\pquo(F, G)$ and $\prem(F, G)$ are used to denote the
\alert{pseudo-quotient} and \alert{pseudo-remainder} of $F$ with respect to
$G$ in $\lv(G)$ respectively.

\begin{definition}
An ordered set
$\pset{T}=[T_1, \ldots, T_r]$ of non-constant polynomials in $\kx$ is called a
\emph{triangular set} if  $\lv(T_i)<\lv(T_j)$ for all $i<j$.
\end{definition}

Suppose that $\pset{T}=[T_1, \ldots, T_r]$ is a triangular set.
We use $y_i$ as an alias of $\lv(T_i)$ for each $i=1,\dots,r$.  Moreover,
$\p{y}_i$ stands for $y_1, \ldots, y_i$ with $\p{y}=\p{y}_r$.
The triangular set $\pset{T}$ is said to be zero-dimensional if $\xvar=\p{y}$.
We denote $\p{u}$ the variables in $\xvar$ {but} not in $\p{y}$.

Let $\ktr$ represent the transcendental extension field $\cnum(\p{u})$. To
avoid ambiguity, for any ideal $\pset{I}\subseteq\cnum[\p{u},\p{y}_i]$,
$\pset{I}_\ktr$ denotes the ideal generated by $\pset{I}$ in
$\ktr[\p{y}_i]$.
The
\alert{saturated ideal} of  $\pset{T}$ is defined as
$$\sat(\pset{T})\equiv \bases{\pset{T}} : H^{\infty}\equiv
\{F:\, \text{there exits an integer}~s~\text{such that}~ FH^s\in\bases{\pset{T}}\},$$
where $H$ is the product of the initials of all polynomials in $\pset{T}$.
Moreover, we define
$\sat_i(\pset{T})\equiv\sat([T_1,\dots,T_i])$.

\begin{definition}
  Let $\pset{T}=[T_1, \ldots, T_r]\subseteq\kx$ be a triangular set. $\pset{T}$
  is called a \alert{regular set} in $\kx$ if for each $i=1,\ldots,r$,
  $\ini(T_i)$ is neither zero nor a zero divisor in quotient ring
  $\kx/\sat_{i-1}(\pset{T})$.
\end{definition}

The notation of regular set was introduced first by Kalkbrener \cite{k93g}, Yang and Zhang \cite{y94s} simultaneously.
In the following, we list two main properties of regular sets.
For more details, readers may refer to \cite{k93g,y94s,h03n,a99t,w00c}.

%
%
%

\begin{proposition}[\cite{h03n}]\label{prop:unmix}
Let $\pset{T}$ be a regular set in $\kx$. Then
  \begin{enumerate}
    \item $\sat(\pset{T})\neq \kx$;
    \item $\pset{T}$ is zero-dimensional if and only if $\sat(\pset{T})$ is a zero-dimensional ideal;
    \item $\sat(\pset{T})$ is an unmixed-dimensional ideal.
  \end{enumerate}
\end{proposition}

\begin{proposition}[\cite{h03n}]\label{prop:satT=T}
  For any regular set $\pset{T}\subseteq\kx$, $\sat(\pset{T})_\ktr =
  \bases{\pset{T}}_\ktr$. Furthermore, $\bases{\pset{T}}_\ktr\cap\kx=\sat(\pset{T})$.
\end{proposition}

Proposition \ref{prop:satT=T} plays a key role in this paper. By this property, we know that
if the regular set $\pset{T}$ is zero-dimensional, then $\sat(\pset{T})= \bases{\pset{T}}$.

Let $F$ be a polynomial
in $\cnum[\p{u},\p{y}_i]$. Then $F$ can also be viewed as an element in
$\ktr[\p{y}_i]$. For any prime ideal $\pset{P}\subseteq
\ktr[\p{y}_{i-1}]$, $\overline{F}^{\pset{P}}$ denotes the image of $F$
in $(\ktr[\p{y}_{i-1}]/\pset{P})[y_i]$ under the natural homomorphism.
For any polynomial set $\pset{S}\in\ktr[\p{y}_i]$, define
$\overline{\pset{S}}^\pset{P}\equiv\{\overline{S}^\pset{P}:\,S\in \pset{S}\}$.

\begin{definition}\label{def:sim}
A regular set $\pset{T}=[T_1, \ldots, T_r]$ in $\kx$ is called a \alert{simple set}
or said to be \alert{simple} if for each $i=1,\ldots, r$ and
associated prime $\pset{P}$ of $\sat_{i-1}(\pset{T})_\ktr$,
$\overline{T}_{i}^{\pset{P}}$ is a squarefree polynomial in
$(\ktr[\p{y}_{i-1}]/\pset{P})[y_i]$.
\end{definition}

{The notion of simple set originates from \cite{t37d,w98d}.}
A similar definition can be found in \cite{h03n},
which is called \emph{squarefree regular chain} therein.
The following proposition reveals the most important property of simple sets.

\begin{proposition}[\cite{l10d}]\label{thm:sqrEquiv}
Let $\pset{T}$ be a regular set in $\kx$. Then
the following statements are equivalent:
\begin{enumerate}
  \item $\pset{T}$ is simple;\label{itm:pro-1}
  \item $\sat(\pset{T})$ is a radical ideal;\label{itm:pro-2}
  \item $\sat(\pset{T})_\ktr$ is a radical ideal.\label{itm:pro-3}
\end{enumerate}
\end{proposition}

\section{Pseudo Squarefree Decomposition Modulo a Regular Set}\label{sec:psqf}

Let $\pset{I}$ and $\pset{I}_1,\dots,\pset{I}_s$ be ideals in $\kx$ with
\begin{equation}\label{eq:irr-decom}
  \pset{I}=\pset{I}_1\cap\cdots\cap\pset{I}_s.
\end{equation}
We say \eqref{eq:irr-decom} is an \emph{irredundant decomposition} if, for any
associated prime $\pset{P}$ of $\pset{I}$, there exists a unique
$i$ such that $\rad{\pset{I}_i}\subseteq \pset{P}$.


\begin{theorem}[\cite{h03n,k93g,m00t,klmsw21}]\label{thm:pgcd}
 There exists an algorithm (named by $\algpgcd$) with  a polynomial set $\pset{F}$ in
$\kx[z]$ and a regular set $\pset{T}$  in $\kx$ as its input, where $\kx[z]$
represents the polynomial ring with all variables in $\p{x}$
smaller than $z$, such that the output
$\{(G_1,\pset{A}_1),\ldots,(G_s,\pset{A}_s)\}$  satisfies the following conditions:
\begin{enumerate}
  \item each $\pset{A}_i$ is a regular set in $\kx$ and
$\sat(\pset{T})\subseteq\sat(\pset{A}_i)$;\label{itm:tsuba}

  \item $\rad{\sat(\pset{T})}=\rad{\sat(\pset{A}_1)}
  \cap\cdots\cap\rad{\sat(\pset{A}_s)}$
  is an irredundant decomposition; \label{itm:spe-dec}

  \item The ideal in
    $\fr(\kx/\sat(\pset{A}_i))[z]$ generated by $\pset{F}$  equals to that generated by the polynomial $G_i$,
    where $\fr(\kx/\sat(\pset{A}_i))$ is the total quotient ring of
$\kx/\sat(\pset{A}_i)$, i.e.\ the localization of $\kx/\sat(\pset{A}_i)$ at
the multiplicatively closed set of all its non-zerodivisors;\label{itm:spe-gcd}
  \item $G_i\in\bases{\pset{F}}+\sat(\pset{A}_i)$;
  \item $G_i=0$, or $\lc(G_i,z)$ is neither zero nor a zero divisor in quotient ring $\fr(\kx/\sat(\pset{A}_i))$.\label{itm:e}
\end{enumerate}
\end{theorem}

\begin{remark}\label{rem:pgcd}
  It is pointed out in \cite{h03n} that if $\pset{T}$ is a simple set, then all $\pset{A}_i$
  in the output of $\algpgcd(\pset{F},\pset{T})$ are also simple sets. Furthermore, the ideal
  relation in \ref{itm:spe-dec} can be replaced with $\sat(\pset{T})=\sat(\pset{A}_1)\cap\cdots\cap\sat(\pset{A}_s)$
  {in this case.}
\end{remark}

 It is known that
$\fr(\kx/\sat(\pset{A}_i))=\ktr[\p{y}]/\sat(\pset{A}_i)_{\ktr}$. Then by \ref{itm:spe-gcd},
for any associated prime $\pset{P}$ of $\sat(\pset{A}_i)_\ktr$, we have that
$\bases{\overline{\pset{F}}^\pset{P}}=\bases{\overline{G}_i^\pset{P}}$, i.e.\
$\gcd(\overline{\pset{F}}^\pset{P})=\overline{G}_i^\pset{P}$.
Therefore, the set $\{(G_1,\pset{A}_1),\ldots,(G_s,\pset{A}_s)\}$ satisfying
the above five conditions  is called the
\alert{pseudo gcd} of $\pset{F}$ modulo $\pset{T}$.

For any univariate polynomials $A$ and $B$, the expression $A\thicksim B$
means that there exists a nonzero constant $c$ such that $A=cB$.
Let $F, A_1,\dots,A_s$ be non-constant polynomial in $\cnum[x]$
and $a_1,\dots,a_s$ be positive integers. We call $\{[A_1,a_1],\ldots,[A_s,a_s]\}$ the
\alert{squarefree decomposition} of $F$ if the following conditions are
satisfied:
\begin{itemize}
  \item $F\thicksim A_1^{a_1}\cdots A_s^{a_s}$,
  \item $\gcd(A_i,A_j)=1$ for all $i\neq j$,
  \item $A_i$ is squarefree for all $i=1,\ldots,s$.
\end{itemize}

The following example illustrates the philosophy of computing
the {squarefree decomposition} of a univariate polynomial.
For relative algorithms, readers may refer to \cite{m03s}.

\begin{example}
Consider the univariate polynomial $F=3\,x^5-3\,x^3\in\cnum[x]$.
First compute $\gcd(F,{\rm d}F/{\rm d}x)$ and store the result in $P$.
It is easy to see that $P=x^2$, which is a factor of $F$. Let $Q=F/P=3\,x^3-3\,x$. Further computing  $\gcd(P,Q)$,
one obtains $x$, which is also a factor of $Q$. Since $Q/x=3\,x^2-3\thicksim x^2-1$,
we have $F\thicksim x^3(x^2-1)$, where  $x$ and $x^2-1$ are coprime {and squarefree}.
As a result, the squarefree decomposition of $F$ is $\{[x,3],[x^2-1,1]\}$.
\end{example}

In \cite{l10d}, the first author of this paper and the coworkers generalized the squarefree decomposition
of a univariate polynomial to the so-called pseudo squarefree
decomposition of a multivariate polynomial modulo a simple set. We slightly modify the definition
of pseudo squarefree
decomposition in \cite{l10d}
as follows.

\begin{definition}\label{def:psqf}
For any regular set $\pset{T}\subseteq \kx$ and polynomial
$F\in \kx[z]\setminus \kx$, the set
$$\{(\{[P_{i1},a_{i1}],\ldots,[P_{ik_i},a_{ik_i}]\},\pset{A}_i):\,i=1,\ldots,s\}$$
is called the \alert{pseudo squarefree
decomposition} of
$F$ modulo $\pset{T}$ if
\begin{enumerate}
\item each $\pset{A}_i$ is a regular set in $\kx$ and
$\sat(\pset{T})\subseteq\sat(\pset{A}_i)$;\label{itm:def-sqf-rad}
\item $\rad{\sat(\pset{T})}=\rad{\sat(\pset{A}_1)}
  \cap\cdots\cap\rad{\sat(\pset{A}_s)}$
    is an irredundant decomposition;\label{itm:def-sqf-dec}
\item each $\{[\overline{P}_{i1}^\pset{P},a_{i1}],\ldots,[\overline{P}_{ik_i}^\pset{P},a_{ik_i}]\}$
is the squarefree decomposition of $\overline{F}^\pset{P}$ for any associated prime
$\pset{P}$ of $\sat(\pset{A}_i)_\ktr$.\label{itm:def-sqf-sqf}
\end{enumerate}
\end{definition}

Moreover, for any $F\in\ffx[z]$ and any zero-dimensional simple set $\pset{T}$ in $\ffx$,
where $\ff$ is a finite field,
an effective algorithm for computing the {pseudo squarefree
decomposition} of $F$ modulo $\pset{T}$ {was} given in \cite{l10d}.
In the sequel, we propose a new algorithm (Algorithm \ref{alg:sqfrx}), obtained by modifying
the algorithm in \cite{l10d}, for computing the {pseudo squarefree
decomposition} of polynomials in $\kx[z]$.

\smallskip
\begin{algorithm}[H]\label{alg:sqfrx}
\caption{Pseudo Squarefree Decomposition $\SQF:=\algpsqf(F,
\pset{T})$}

\KwIn{a polynomial \underline{$F$} in $\kx[z]\setminus\kx$;
a regular set \underline{$\pset{T}$} in $\kx$.}

\KwOut{the pseudo squarefree decomposition \underline{$\SQF$} of $F$ modulo
$\pset{T}$.}

\BlankLine

$\SQF:=\emptyset$; $\TD:=\emptyset$\;

\For{$(C_1,\pset{C})\in \algpgcd(\{F,
\partial F/\partial z\},\pset{T})$\label{line:c1rt} }{

$B_1:=\pquo(F,C_1)$\;\label{line:psqf-B1}

$\TD:=\TD\cup\{[B_1,C_1,\pset{C},\emptyset,1]\}$\;

}\label{line:endeach}

\While{$\TD\neq \emptyset$}{ $[B_1,C_1,\pset{C},\PS,d]:=\pop(\TD)$\;
\label{line:pop}

\eIf{$\deg(B_1,z)>0$}{\For{$(B_2,\pset{A})\in
\algpgcd(\{B_1,C_1\},\pset{C})$\label{line:eachpgcd}}{
$C_2:=\pquo(C_1,B_2)$\;\label{line:psqf-C2}
$P:=\pquo(B_1,B_2)$\;\label{line:psqf-P} \lIf{$\deg(P,z)>0$}
{$\PS:=\PS\cup\{[P,d]\}$\;}
$\TD:=\TD \cup \{[B_2,C_2,\pset{A},\PS,d+1]\}$\;\label{line:son}}
}{
$\SQF:=\SQF \cup \{(\PS,\pset{C})\}$\;
}

}

$\alg{return}(\SQF)$\;
\end{algorithm}

We use $\pop(\TD)$ to represent
the operation of taking one element randomly and then delete it from $\TD$.
In Algorithm \ref{alg:sqfrx}, $\TD$ stores what to be processed. For each element
$[B,C,\pset{C},\PS,d]\in\TD$, one may see that $\pset{C}$ is a regular set over which
later computation is to be performed, and  $\PS$ stores the squarefree
components already obtained with exponent smaller than $d$.

It can be observed that the {\bf while} loop  is essentially a
splitting procedure. Thus we may regard the running of Algorithm \ref{alg:sqfrx}  as building trees with
elements in $\TD$ as their nodes. The roots of these trees are constructed in the first {\bf for} loop.
For each node
$[B_1,C_1,\pset{C},\PS,d]$, its child $[B_2,C_2,\pset{A},\PS,d+1]$
is built when the statement ``$\TD:=\TD \cup \{[B_2,C_2,\pset{A},\PS,d+1]\}$'' is executed.
For any fixed path of one of the trees, we denote the node of depth
$i$ in the path by $[B{(i)},C{(i)},\pset{C}{(i)},\PS{(i)},i]$.

{\noindent \textbf{Correctness.}
The conditions \ref{itm:def-sqf-rad} and
\ref{itm:def-sqf-dec} of Definition \ref{def:psqf} follow from
Theorem \ref{thm:pgcd} \ref{itm:tsuba} and  \ref{itm:spe-dec} respectively.

To prove Theorem \ref{thm:pgcd} \ref{itm:def-sqf-sqf}, the tool of localization may be helpful.
Suppose that $[B{(s)},C{(s)},\pset{C}{(s)},\PS{(s)},t]$ is a leaf node of the tree.
For any associated prime $\pset{P}$
of $\sat(\pset{C}{(t)})_{\ktr}$, $\overline{F}^\pset{P}$ is a
univariate polynomial over the field $\ktr[\p{y}]/\pset{P}$.
We can assume that $\overline{F}^\pset{P}=\prod_{i=1}^{t} P_i^i$,
where $P_i$ are squarefree polynomial in $z$ and $\gcd(P_j,P_k)=1$ for any $j\neq k$.
It can be proved that
\begin{equation}\label{eq:bc}
\overline{B{(i)}}^\pset{P}\thicksim P_iP_{i+1}\cdots
P_{t}\qquad\text{and}\qquad\overline{C{(i)}}^\pset{P}\thicksim
P_{i+1}P_{i+2}^2\cdots P_{t}^{t-i}.
\end{equation}
Thus $\overline{B{(i)}}^\pset{P}/\overline{B{(i-1)}}^\pset{P}=P_i$.
Therefore $\PS$ stores the squarefree decomposition of $\overline{F}^\pset{P}$.\hfill$\square$
}

{\noindent\textbf{Termination.}
It suffices to prove that every path in the tree is finite, which is obvious by
\eqref{eq:bc}.\hfill$\square$
}

%

\section{Analyzing Multiplicity}\label{sec:mult}

In this section, we propose algorithms for analyzing multiplicity of a regular set's zeros.
{As a preparation, the following algorithm is given first, which can be used to  decompose any given regular
  set over $\cnum$ into a finite number of simple sets.}

\smallskip
\begin{algorithm}[H]\label{alg:simdec}
\caption{$\mathbb{S}:=\alg{Reg2Sim}(\pset{T})$}

\KwIn{a regular set \underline{$\pset{T}$} in $\kx$.}

\KwOut{a finite set \underline{$\mathbb{S}$} with elements of the form $(\pset{B},P)$, where $\pset{B}=[B_1,\dots,B_r]$ is
a simple set in $\kx$ and $P=[p_1,\dots,p_r]$ is an array of integers. We use $\pset{B}^P$ to denote $[B_1^{p_1},\dots,B_r^{p_r}]$
and call $P$ the \emph{multiplicity array} of $\pset{B}^P$.
Furthermore, we have that
\begin{equation}\label{eq:reg2sim}
\sat(\pset{T})=\bigcap_{(\pset{B},P)\in
\mathbb{S}}\sat(\pset{B}^P),
\end{equation}
which is an irredundant decomposition.
}

\BlankLine

$\mathbb{S}:=\emptyset$; $\TD:=\{(\ts,[\,],[\,])\}$\;

 \While{$\TD\neq \emptyset$} {
    $(\pset{A}, \pset{B}, P):=\pop(\TD)$\;
    \eIf{$\pset{A}=\emptyset$}{
        $\mathbb{S}:= \mathbb{S}\cup \{(\pset{B},P)\}$\;
        \label{line:addB}
    }{
    $A:=$ the first polynomial in $\pset{A}$\;
    \For{$(\{[C_1,c_1],\ldots,[C_s,c_s]\},\pset{Q})\in \algpsqf(A,\pset{B})$\label{line:zong-for-1}}{

        $\TD:=\bigcup_{i=1}^s\{(\pset{A}\setminus\{A\}, \alg{append}(\pset{Q},C_i), \alg{append}(P,c_i))\}\cup \TD$\;\label{line:wdec-td}
    }\label{line:zong-for-2}
    }
}

$\alg{return}(\mathbb{S})$\;
\end{algorithm}

In Algorithm \ref{alg:simdec}, $\alg{append}(L,a)$ returns the array obtained
by appending the element $a$ to the end of $L$.
The termination is obvious. In order to prove the correctness, the following lemma
is needed.

\begin{lemma}\label{lem:split}
 Suppose that  $\pset{T}$ is a simple set in $\kx$ and $F$ is a  polynomial in
$\kx[z]\setminus \kx$. Let
$\{(\{[P_{i1},a_{i1}],\ldots,[P_{ik_i},a_{ik_i}]\},\pset{A}_i):\,i=1,\ldots,s\}$ be
the output of {\rm$\alg{Reg2Sim}(F,\pset{T})$}. Then all $\pset{A}_i$ are simple sets.
Furthermore,  $\sat(\pset{T})=\sat(\pset{A}_1)\cap\cdots\cap\sat(\pset{A}_s)$.
\end{lemma}

\begin{proof}
It directly follows from Remark \ref{rem:pgcd}.
\end{proof}

%

\noindent\textbf{Correctness}(Algorithm \ref{alg:simdec})\textbf{.}
For any element $(\pset{B},P)$ in the output of $\alg{Reg2Sim}(\pset{T})$, one can easily know that
$\pset{B}$ is a simple set by Lemma \ref{lem:split} and Definition \ref{def:sim}.

The ideal relation \eqref{eq:reg2sim} could be proved as follows. For each $(\pset{A},\pset{B},P)\in\TD$ which satisfies that
$\pset{A}\neq\emptyset$, the statement ``$(\{[C_1,c_1],\ldots,[C_s,c_s]\},\pset{Q})\in \algpsqf(A,\pset{B})$''
in the {\bf for} loop is then executed. It can be observed that
$$\bases{\pset{A}}_\ktr+\bases{\pset{B}^P}_\ktr=
\bigcap_{(\{[C_1,c_1],\ldots,[C_s,c_s]\},\pset{Q})\in \algpsqf(A,\pset{B})}\bases{\pset{A}}_\ktr+\bases{\pset{Q}^P}_\ktr.$$
Furthermore, for each $(\{[C_1,c_1],\ldots,[C_s,c_s]\},\pset{Q})\in \algpsqf(A,\pset{B})$, we have that
\begin{equation*}
  \begin{split}
    \bases{\pset{A}}_\ktr+\bases{\pset{Q}^P}_\ktr&=\bases{\pset{A}\setminus\{A\}}_\ktr+\bases{\pset{Q}^P\cup\{A\}}_\ktr\\
    &=\bases{\pset{A}\setminus\{A\}}_\ktr+ \bases{\pset{Q}^P\cup\{\prod_{i=1}^s C_i^{c_i}\}}_\ktr \\
    &=\bases{\pset{A}\setminus\{A\}}_\ktr+ \bigcap_{i=1}^s \bases{\pset{Q}^P\cup\{C_i^{c_i}\}}_\ktr\\
    &=\bigcap_{i=1}^s \bases{\pset{A}\setminus\{A\}}_\ktr+ \bases{\pset{Q}^P\cup\{C_i^{c_i}\}}_\ktr.
  \end{split}
\end{equation*}
Thus in the {\bf while} loop, we have the following invariant:
$$\bases{\pset{T}}_\ktr=\bigcap_{(\pset{A},\pset{B},P)\in\TD}\bases{\pset{A}\cup\pset{B}^P}_\ktr
\cap\bigcap_{(\pset{B},P)\in\mathbb{S}}\bases{\pset{B}^P}_\ktr.$$
When the {\bf while} loop terminates, $\bases{\pset{T}}_\ktr=\bigcap_{(\pset{B},P)\in\mathbb{S}}\bases{\pset{B}^P}_\ktr$.
Intersecting the left and right sides of this equation with $\kx$, we obtain  \eqref{eq:reg2sim}.

The irredundant property of the ideal decomposition in \eqref{eq:reg2sim}
follows from Definition \ref{def:psqf} and the property of Algorithm \ref{alg:sqfrx}.\hfill$\square$

In what follows, we show how the multiplicity arrays in the output of $\alg{Reg2Sim}(\pset{T})$ are used
to count the multiplicities at zeros of $\pset{T}$.

\begin{lemma}\label{lem:neq}
Suppose that $[B_1,\dots,B_r]$ is a zero-dimensional simple set in $\kx$, and $[p_1,\dots,p_r]$ is a list of integers.
Let $\pp=(a_1,\ldots,a_r)$ be a zero of $\pset{I}=\sat([B_1,\dots,B_r])$ and $\partial_{j_1\cdots j_r}$ be
a differential functional with $j_i\geq p_i$ for some $i$'s. Then there exists a polynomial $F_{j_1\cdots j_r}$ in $\pset{I}$
such that $\partial_{j_1\cdots j_r}[\pp](F_{j_1\cdots j_r})\neq 0$.
\end{lemma}
\begin{proof}
  Suppose that $\mu$ is the smallest integer among $i$'s such that $j_i\geq p_i$.
  Let
  $$F_{j_1\cdots j_r}=\Big(\prod_{k\neq\mu}(x_k-a_k)^{j_k}\Big)(x_\mu-a_\mu)^{j_\mu-p_\mu}B_\mu^{p_\mu}.$$
  It is obvious that $F_{j_1\cdots j_r}\in\pset{I}$.
  For any polynomial $P\in\kx$,
  \begin{equation}\label{eq:diff}
    \partial_{j_1\cdots j_r}[\pp](P)=\partial_{j_\mu}\Big(\partial_{j_1\cdots j_{\mu-1}j_{\mu+1}
  \cdots j_r}(P)|_{(a_1,\dots,a_{\mu-1},a_{\mu+1},\dots,a_r)}\Big)|_{x_\mu=a_\mu}.
  \end{equation}
  Denote $\partial_{j_1\cdots j_{\mu-1}j_{\mu+1}
  \cdots j_r}(F_{j_1\cdots j_r})|_{(a_1,\dots,a_{\mu-1},a_{\mu+1},\dots,a_r)}$ by $G$.
  Since  $$\partial_{j_1\cdots j_{\mu-1}j_{\mu+1}\cdots j_r}\Big(\prod_{k\neq\mu}(x_k-a_k)^{j_k}\Big)=1,$$
  we have
  \begin{equation*}
 G=(x_\mu-a_\mu)^{j_\mu-p_\mu}\Big(B_\mu|_{(a_1,\dots,a_{\mu-1},a_{\mu+1},\dots,a_r)}\Big)^{p_\mu},
  \end{equation*}
  where $B_\mu|_{(a_1,\dots,a_{\mu-1},a_{\mu+1},\dots,a_r)}$ is a squarefree polynomial in $\cnum[x_\mu]$.
  It is known that $(a_1,\ldots,a_r)$ is a zero of $\pset{I}$ and $B_\mu\in\pset{I}$, thus one can assume that
  $$B_\mu|_{(a_1,\dots,a_{\mu-1},a_{\mu+1},\dots,a_r)}=(x_\mu-a_\mu)\cdot A,$$
  where $A\in\cnum[x_\mu]$ and $\gcd(x_\mu-a_\mu,A)=1$. Therefore $G=(x_\mu-a_\mu)^{j_\mu}A^{p_\mu}$.
   By \eqref{eq:diff},
  $$\partial_{j_1\cdots j_r}[\pp](F_{j_1\cdots j_r})=\partial_{j_\mu}(G)|_{x_\mu=a_\mu}=A^{p_\mu}|_{x_\mu=a_\mu}\neq 0,$$
  which completes the proof.
\end{proof}

\begin{proposition}\label{prop:bas}
  Let $\pset{B}=[B_1,\dots,B_r]$ be a zero-dimensional simple set in $\kx$, and $P=[p_1,\dots,p_r]$ be a list of integers.
  Then for any zero  $\pp=(a_1,\ldots,a_r)$ of $\pset{I}=\sat(\pset{B}^P)$, the dual space $\mathbb{D}_{\pp}(\pset{I})$
  is spanned by
  $$S=\Big\{\partial_{j_1\cdots j_r}[\pp]:\,0\leq j_i<p_i~\text{for all}~i=1,\dots,r\Big\}.$$
\end{proposition}

\begin{proof}
  Suppose that  $\partial_{j_1\cdots j_r}$ satisfies that $0\leq j_i<p_i$ for all $i=1,\dots,r$.
  It is easy to verify that $B_i\,|\,\partial_{j_1\cdots j_r}(B_i^{p_i})$,
  i.e.\ there exists a polynomial $A_i\in\kx$ such that $\partial_{j_1\cdots j_r}(B_i^{p_i})=A_iB_i$.
  Since $\pset{B}^P$ is a zero-dimensional regular set, we know that
  $$\sat(\pset{B}^P)=\bases{B_1^{p_1},\dots,B_r^{p_r}}.$$
  For any $F\in\sat(\pset{B}^P)$, there exist $C_1,\dots,C_r\in\kx$ such that
  $F=\sum_{i=1}^rC_iB_i^{p_i}.$
  Thus $$\partial_{j_1\cdots j_r}(F)=\sum_{i=1}^r B_i[ \partial_{j_1\cdots j_r}(C_i)B_i^{p_i-1}+A_iC_i].$$
  Since $B_i(\pp)=0$, it follows that $\partial_{j_1\cdots j_r}[\pp](F)=0$.
  Hence $\partial_{j_1\cdots j_r}[\pp]\in\mathbb{D}_{\pp}(\pset{I})$.

  On the other hand, suppose that $\sum_{i=1}^l c_{\js_i}\partial_{\js_i}[\pp]$~($c_{\js_i}\in\cnum$)
  is a differential functional in $\mathbb{D}_{\pp}(\pset{I})$.
  Without loss of generality, one may assume that $\partial_{\js_i}[\pp]\not\in S$
  for $i=1,\ldots,m$ and $\partial_{\js_i}[\pp]\in S$ for $i=m+1,\ldots,l$.
  For each $\partial_{\js_k}[\pp]$, $k=1,\ldots,m$, construct $F_{\js_k}\in\pset{I}$
  such that $\partial_{\js_k}[\pp](F_{\js_k})\neq 0$
  in the same way as we did in the proof of Lemma \ref{lem:neq}.
  For any $i=m+1,\ldots,l$, it can be proved that $\partial_{\js_i}[\pp](F_{\js_k})=0$.
  Furthermore,  $\partial_{\js_i}[\pp](F_{\js_k})=0$ if
  $i=1,\dots,m$ and $i\neq k$. It follows that
  $$\sum_{i=1}^l c_{\js_i}\partial_{\js_i}[\pp](F_{\js_k})=c_{\js_k}\partial_{\js_k}[\pp](F_{\js_k})=0,
  ~\text{for}~k=1,\dots,m.$$
  Since $\partial_{\js_k}[\pp](F_{\js_k})\neq0$, we know that $c_{\js_1}=\cdots=c_{\js_m}=0$,
  which means that $\mathbb{D}_{\pp}(\pset{I})$ is spanned by $S$.
  The proof is complete.
\end{proof}

\begin{corollary}\label{cor:mult}
  Let $\pset{B}=[B_1,\dots,B_r]$ be a zero-dimensional simple set in $\kx$, and $P=[p_1,\dots,p_r]$ be a list of integers.
  Then the local multiplicity of any zero in $\sat(\pset{B}^P)$ is $\prod_{i=1}^r p_i$.
\end{corollary}
\begin{proof}
  This is obvious by Definition \ref{def:dual-mult} and Proposition \ref{prop:bas}.
\end{proof}


%

The following lemma states a classical result in commutative algebra
(e.g., see \cite{a69i} for its proof).

\begin{lemma}\label{lem:froma69i}
Suppose that $\pset{S}$ is a multiplicatively closed subset of $\kx$, and $\pset{I}$, $\pset{I}_1$, $\pset{I}_2$ are
polynomials in $\kx$.
\begin{enumerate}
 \item $\pset{S}^{-1}(\pset{I}_1\cap\pset{I}_2)=\pset{S}^{-1}\pset{I}_1\cap\pset{S}^{-1}\pset{I}_2$.
  \item If $\pset{S}\cap\pset{P}\neq\emptyset$ for every prime ideal $\pset{P}\supseteq\pset{I}$,
  then $\pset{S}^{-1}\pset{I}=\pset{S}^{-1}\kx$.
\end{enumerate}
\end{lemma}

\begin{theorem}[Main Theorem]\label{thm:main}
Suppose that a zero-dimensional regular set $\pset{T}\subseteq\kx$ is given. Let
$\mathbb{S}=\{(\pset{B}_1,P_1),\dots,(\pset{B}_k,P_k)\}$ be the output of {\rm$\alg{Reg2Sim}(\pset{T})$}.
For any zero $\pp=(a_1,\dots,a_r)$ of $\bases{\pset{T}}$, there exists one and only one element
$([B_1,\dots,B_r],[p_1,\dots,p_r])\in\mathbb{S}$ such that $B_1(\pp)=0,\dots,B_r(\pp)=0$.
Furthermore, the local multiplicity of $\pp$ in $\bases{\pset{T}}$ is $\prod_{i=1}^r p_i$.
\end{theorem}
\begin{proof}
  The existence of such $([B_1,\dots,B_r],[p_1,\dots,p_r])\in\mathbb{S}$ is from
   \eqref{eq:reg2sim}. While the uniqueness is because the decomposition in \eqref{eq:reg2sim}
  is irredundant.

  Without loss of generality, we assume that
  $$\{(\pset{B}_1,P_1),\dots,(\pset{B}_k,P_k)\}=\alg{Reg2Sim}(\pset{T})$$
  with $(\pset{B}_1,P_1)=([B_1,\dots,B_r],[p_1,\dots,p_r])$.
  By Theorem \ref{def:inter-mult} and Corollary \ref{cor:mult}, it suffices to prove
  $\pset{S}^{-1}\bases{\pset{T}}=\pset{S}^{-1}\sat(\pset{B}_1^{P_1})$,
  where  $\pset{S}=\kx\setminus\pset{M}_\pp$ and $\pset{M}_\pp=\bases{x_1-a_1,\ldots,x_r-a_r}$.

  We know that $\bases{\tss}=\sat(\tss)$.
  By Lemma \ref{lem:froma69i} (i) and \eqref{eq:reg2sim},
  $$\pset{S}^{-1}\bases{\pset{T}}=\pset{S}^{-1}\sat(\pset{T})=\bigcap_{i=1}^k \pset{S}^{-1}\sat(\pset{B}_i^{P_i}).$$
  As $B_1(\pp)=0,\dots,B_r(\pp)=0$, we have that $\sat(\pset{B}_1^{P_1})=\bases{\pset{B}_1^{P_1}}\subseteq \pset{M}_\pp$.
  Moreover, \eqref{eq:reg2sim} is an irredundant
  decomposition, thus $\sat(\pset{B}_i^{P_i})\not\subseteq \pset{M}_\pp$ for any $i\neq 1$.
  Then it can be proved that $\pset{S}\cap\pset{P}\neq\emptyset$
  for every prime ideal $\pset{P}\supseteq\sat(\pset{B}_i^{P_i})$, $i\neq1$.
  By Lemma \ref{lem:froma69i} (ii), $\pset{S}^{-1}\sat(\pset{B}_i^{P_i})=\pset{S}^{-1}\kx$ for any $i\neq 1$.
  Hence $\pset{S}^{-1}\bases{\pset{T}}=\pset{S}^{-1}\sat(\pset{B}_1^{P_1})$.
\end{proof}

{By the above theorem, one can easily count the multiplicities at zeros of any given
zero-dimensional regular set $\tss$  from the output of $\alg{Reg2Sim}(\tss)$.
The following example illustrates the idea.}

\begin{example}\label{ex:mult}
Consider the following regular set in $\cnum[x,y]$:
\begin{equation*}
 \tss=[x^3-x^2+2,(x^5+x)y^3-x^3y^2].
\end{equation*}
Applying $\alg{Reg2Sim}$ to $\tss$, we obtain the output of $4$ branches:
\begin{equation*}
  \begin{split}
    &(\pset{B}_1,P_1)=([ x^2  - 2\, x + 2, y],~ [1, 2]),\\
    &(\pset{B}_2,P_2)=([ x + 1, 2\, y-1,],~ [1, 1]),\\
    &(\pset{B}_3,P_3)=([x + 1,y], ~[1, 2]),\\
    &(\pset{B}_4,P_4)=([x ^2 - 2\, x + 2, (3\, x - 3) y-2],~[1,1]).
  \end{split}
\end{equation*}
To count the multiplicity at, e.g., the complex zero $\pp=(1+{\rm i}, 0)$ of $\tss$,
one just check that $\pp$ is a zero of $\pset{B}_1$. Then from  $P_1$, we
know that the multiplicity of $\pp$ is $2$.
\end{example}

{We give a description of the input and output of the function for computing
the multiplicity as follows without entering the details.}

\smallskip
\begin{algorithm}[H]
\caption{$M:=\alg{RegMult}(\pset{T},\pp)$}

\KwIn{a zero-dimensional regular set \underline{$\pset{T}$} in $\kx$;
a zero \underline{$\pp$} of $\tss$.}

\KwOut{the local multiplicity of $\pp$ in $\sat(\tss)$.
}
\end{algorithm}

It should be noted that $\alg{Reg2Sim}$ computes not the multiplicity of just one zero of a regular set,
but {essentially} the multiplicities of all its zeros.

\begin{remark}
The multiplicity array produced by {\rm$\alg{Reg2Sim}$} may
be more appropriate than {the local multiplicity in} Definition \ref{def:dual-mult}
for  characterizing the multiplicity.
For example, consider ideals
$\bases{x^2,y^3}$ and $\bases{x^3,y^2}$ in $\cnum[x,y]$. It is easy to see that $(0,0)$
is their {unique} zero, and the local multiplicities of $(0,0)$
in these two ideal both equal to $6$. But it is obvious that $\bases{x^2,y^3}\neq\bases{x^3,y^2}$,
and their \grobner bases are different under a same term order.

It is well known that the \grobner basis is one of elimination methods that preserve the multiplicity.
From the above example, we know that the multiplicity in the \grobner sense differs from
 the local multiplicity, but is closer to the multiplicity array.
 For the above example, the multiplicity array $[2,3]$  of $\bases{x^2,y^3}$
 is distinct from   the multiplicity array $[3,2]$ of $\bases{x^3,y^2}$.
 It never occurs that ideals of zero-dimensional regular sets are different
  but with same zeros and same multiplicity arrays.
\end{remark}

\begin{remark}
 In \cite{z11r}, Zhang and others proposed an approach for isolating real solutions of a
zero-dimensional triangular set as well as counting their multiplicities.
Interested readers may refer to \cite{z11r} for the formal notation of \emph{real solution isolation with multiplicity}.
It should be noted that
the real solution isolation with multiplicity of any given zero-dimensional regular set $\tss$
can also be easily obtained from the output of $\alg{Reg2Sim}(\tss)$.

The first step is to compute the real solution isolation of $\tss$, i.e.\
``boxes" of the form $\big[[a_1,b_1],\dots,[a_n,b_n]\big]$ with rational $a_i$ and $b_i$
such that each box contains exact one real zero of $\tss$, which could be done by applying,
e.g., the method of \cite{x02a}.
Let $[\pset{B}_1,P_1],\dots,[\pset{B}_k,P_k]$ be the output of $\alg{Reg2Sim}(\pset{T})$.
For each box $\big[[a_1,b_1],\dots,[a_n,b_n]\big]$ that covers one zero (say $\pp$) of $\tss$,
we just need to find the unique $\pset{B}_i$ with $\pp$ as its zero using, e.g.,
the algorithms in \cite{x02c}. Then the multiplicity of $\pp$ can be directly read from $P_i$.

The approach by Zhang and others \cite{z11r}
computes the simple decomposition of a zero-dimensional
generic triangular set  with respect to its real solutions and multiplicities.
On the other hand, ours can produce the simple decomposition
of a zero-dimensional regular set with respect to  all its solutions and multiplicities.
However, the former method needs not to split triangular sets
in the squarefree decomposition over algebraic extension fields, thus may
be more efficient.
\end{remark}

\section{Experimental Results}\label{sec:exp}

{Based on the RegularChains
library in Maple 13, we have implemented the algorithms proposed in this paper}. The Maple package Apatools \cite{z08a} also
provides us with a function {for computing the multiplicity of a zero at any zero-dimensional ideal:}
$${\alg{MultiplicityStructure}(\text{idealBases, variables, zero, threshold}),}$$
which is built on the dual space theory and
can be executed  symbolically or approximately.
In order to be fair, we compare our implementation  with
the symbolic version of $\alg{MultiplicityStructure}$ {by setting the parameter ``threshold'' to be $0$.}

All the experiments were running on a laptop with Intel Core i3-2350TM CPU 2.30 GHz, 2G RAM
and Windows 7 OS. Table 1 records the timings of selected examples,
which are listed in the appendix.

\begin{table}[!h]
\newcolumntype{.}{D{.}{.}{-1}}
\label{tab:time} \caption{Timings of $\alg{MultiplicityStructure}$ and $\alg{RegMult}$ (in seconds)} \centering
\begin{tabular}{c c c c c c}
\hline\hline
No. & Variables & Zero & Multiplicity  & $\alg{MultiplicityStructure}$ & $\alg{RegMult}$\\

\hline
$\tss_1$ & $[x,y]$ & (1,1)  &  1  & .109 &  .093 \\
$\tss_2$ & $[x,y]$ & (1,1)  &  20  & 41.840 &  .047  \\
$\tss_3$ & $[x,y]$ & (2,1)  &  50  & 10.593 &  240.990  \\
$\tss_4$ & $[x,y]$ & (2,1)  &  105  & 120.932 &  3.057 \\
$\tss_5$ & $[u,s]$ & (0,0)  &  6  & 0.187 &  .078 \\
$\tss_6$ & $[u,s,t,x,y,z]$ & (0,0,0,0,0,0)  &  6  & out of memory &  .266 \\
$\tss_7$ & $[x,y,z]$ & (0,0,0)  &  18 & 2.606 &  .046 \\
$\tss_8$ & $[u,s,t,x,y,z]$ & (0,0,0,0,0,0)  &  18  & out of memory &  .172 \\
$\tss_9$ & $[u,s,t,x,y,z]$ & (0,0,0,0,0,0)  &  4  & 34.383 &  1.263 \\
$\tss_{10}$ & $[u,s,t,x,y,z]$ & (0,0,0,0,0,0)  &  24  & out of memory &  1.076 \\
\hline\hline
\end{tabular}
\end{table}

From Table 1, we can observe that $\alg{RegMult}$ is much more efficient
than $\alg{MultiplicityStructure}$ in most cases except $\tss_3$.
One possible reason of the low efficiency of our method on $\tss_3$ is
that the computation of $\alg{psqf}(F,\pset{T})$ may be quite heavy
if the regular set $\pset{T}$ is complex and the factors of $F$ have high exponents.

One can also see that the efficiency of $\alg{MultiplicityStructure}$  decreases
rapidly with the multiplicity going up. Moreover, if the number of variables
is big, the multiplicity matrix (the most important intermediate object in the execution of $\alg{MultiplicityStructure}$)
may become huge even though the involved regular set has simple structure.
In this case, the computation of $\alg{MultiplicityStructure}$ could be fairly time-consuming and the needed memory space
would be unimaginable.
However, our new algorithms do not suffer from these problems.


\section*{Acknowledgements}

{
The authors wish to thank Dongming Wang and Bican Xia for beneficial discussions.

This work has been supported by National Natural Science Foundation of China (No.\ 11601023).
}

\bibliographystyle{elsarticle-num}
\bibliography{multsqf}

\section*{Appendix: Examples in Timings}
\noindent
$\tss_1=[x(x-1),y^{20}(y-1)]$.

\noindent
$\tss_2=[x(x-1)^{20},y(y-1)]$.

\noindent
$\tss_3=[1235556(x-2)^5(234156\,x^4+3456\,x+23677134)^2,
23566234(x^3+23\,x)(y-1)^{10}(x^2y^3+\\~\hspace{25pt}2346234\,y) ]$.

\noindent
$\tss_4=[1235556(x-2)^{21}(234156\,x^4+3456\,x+23677134)^2,
23566234(x^3+23\,x)(y-1)^5(x^2y^3+\\~\hspace{25pt}2346234\,y)]$.

\noindent
$\tss_5=[u^2(u-1)(u^2+u+1),
((u+1)s^3-u)(s^4+1)]$.

\noindent
$\tss_6=[u^2(u-1)(u^2+u+1),
((u+1)s^3-u)(s^4+1),t,x,y,z]$.

\noindent
$\tss_7=[1275467\,x^3(23564882\,x-60289123),
2892349145(y-x)^2(912318912759\,y+29375\,x-\\~\hspace{25pt}12366),(7987326611\,z^2-9712375656\,xy^2)z]$.

\noindent
$\tss_8=[u,s,t,
1275467\,x^3(23564882\,x-60289123),
2892349145(y-x)^2(912318912759\,y+\\~\hspace{25pt}29375\,x-12366),(7987326611\,z^2-9712375656\,xy^2)z]$.

\noindent
$\tss_9=[u(u-1),
(s-u)(s+u-1),
(t-s)(t+u+s-1),
(x-t)(x+u+s+t-1),(y-\\~\hspace{25pt}x)(y+u+s+t+x-1),
(z-y)^4(z+u+s+t+x+y-1)]$.

\noindent
$\tss_{10}=[u^2(u-1),
(s-u)(s+u-1),
(t-s)^2(t+u+s-1),
(x-t)^3(x+u+s+t-1),(y-\\~\hspace{25pt}x)^2(y+u+s+t+x-1),
(z-y)(z+u+s+t+x+y-1)]$.

\end{document}

%% file: macro.tex
\newcommand{\xvar}{\bm{x}}

\newcommand{\js}{{\bm{j}}}

\newcommand{\pp}{{\bm{a}}}

\newtheorem{definition}{Definition}[section] 

 \newtheorem{corollary}{Corollary}[section]
 \newtheorem{lemma}{Lemma}[section]
 \newtheorem{proposition}{Proposition}[section]
 \newtheorem{example}{Example}[section]

 \newtheorem{theorem}{Theorem}[section]
 \newtheorem{remark}{Remark}[section]

\DeclareMathOperator{\fr}{fr}

\newcommand{\ktr}{{\tilde{\cnum}}}

\newcommand{\ff}{{\mathbb{F}_q}}
\newcommand{\ffx}{{\mathbb{F}_q}[\point{x}]}

\newcommand{\alg}[1]{\textsf{#1}}

\DeclareMathOperator{\algpsqf}{\sf psqf}

\DeclareMathOperator{\algpgcd}{\sf pgcd}

\DeclareMathOperator{\pop}{\sf pop}

\newcommand{\SQF}{\mathbb{S}}

\newcommand{\TD}{\mathbb{D}}

\newcommand{\PS}{\mathbb{P}}

\newcommand{\ts}{\pset{T}}

\DeclareMathOperator{\ini}{ini}

\DeclareMathOperator{\lc}{lc}

\DeclareMathOperator{\lv}{lv}

\DeclareMathOperator{\pquo}{pquo}
\DeclareMathOperator{\prem}{prem}
\DeclareMathOperator{\sat}{sat}

\newcommand{\alert}[1]{\emph{#1}}

\newcommand{\pset}[1]{\mathcal{#1}}

\newcommand{\bases}[1]{\langle #1 \rangle}
\newcommand{\point}[1]{\bm{#1}}
\newcommand{\p}[1]{\point{#1}}

\newcommand{\nnum}{\mathbb{N}}

\newcommand{\cnum}{\mathbb{C}}

\newcommand{\kxring}{\cnum[\point{x}]}

\newcommand{\kx}{\kxring}

\newcommand{\rad}[1]{\sqrt{#1}}

\newcommand{\grobner}{Gr\"{o}bner }

\newcommand{\rli}[1]{}

\newcommand{\tss}{\pset{T}}